\documentstyle[12pt]{article}
\begin{document}
\newcommand{\ket}[1] {\mbox{$ \vert #1 \rangle $}}
\newcommand{\bra}[1] {\mbox{$ \langle #1 \vert $}}
\newcommand{\bkn}[1] {\mbox{$ < #1 > $}}
\newcommand{\bk}[1] {\mbox{$ \langle #1 \rangle $}}
\newcommand{\scal}[2]{\mbox{$ \langle #1 \vert #2  \rangle $}}
\newcommand{\expect}[3] {\mbox{$ \bra{#1} #2 \ket{#3} $}}
\newcommand{\ki}{\mbox{$ \ket{\psi_i} $}}
\newcommand{\bi}{\mbox{$ \bra{\psi_i} $}}
\newcommand{\p} \prime
\newcommand{\e} \epsilon
\newcommand{\la} \lambda
\newcommand{\om} \omega   \newcommand{\Om} \Omega
\newcommand{\cc}{\mbox{$\cal C $}}
\newcommand{\w} {\hbox{ weak }}
\newcommand{\al} \alpha
\newcommand{\bt} \beta

\newcommand{\be} {\begin{equation}}
\newcommand{\ee} {\end{equation}}
\newcommand{\ba} {\begin{eqnarray}}
\newcommand{\ea} {\end{eqnarray}}

\def\lrD{\mathrel{{\cal D}\kern-1.em\raise1.75ex
\hbox{$\leftrightarrow$}}}

\def\lr #1{\mathrel{#1\kern-1.25em\raise1.75ex
\hbox{$\leftrightarrow$}}}

\overfullrule=0pt \def\sqr#1#2{{\vcenter{\vbox{\hrule height.#2pt
          \hbox{\vrule width.#2pt height#1pt \kern#1pt
           \vrule width.#2pt}
           \hrule height.#2pt}}}}
\def\square{\mathchoice\sqr68\sqr68\sqr{4.2}6\sqr{3}6}
\def\lrpartial{\mathrel
{\partial\kern-.75em\raise1.75ex\hbox{$\leftrightarrow$}}}

\begin{flushright}
10th December 1998
\\
\end{flushright}
\vskip 1.5 truecm
\centerline{\LARGE\bf{TIME IN COSMOLOGY}}
\vspace{1cm}
\vskip .5 truecm
\centerline{{\bf
R. Brout}}
\centerline{ Service de Physique Th\'eorique}
\centerline{ Universit\'e Libre de Bruxelles, Campus Plaine, CP 225 }
\centerline{ Boulevard du Triomphe, B-1050 Bruxelles, Belgium} 
\centerline{E-mail: rbrout@ulb.ac.be}
\medskip 
\medskip 
\medskip 
\centerline{{\bf R. Parentani}}
\centerline{Laboratoire de Math\'ematiques et Physique Th\'eorique,
CNRS UPRES A 6083,}
\centerline{Facult\' e des Sciences, Universit\'e de Tours, 37200 Tours, France. }
\centerline{E-mail: parenta@celfi.phys.univ-tours.fr}
\vskip 1.5 truecm
\vskip 1. truecm

{\bf Abstract }
The notion of time in cosmology is revealed through an 
examination of transition matrix elements of radiative processes
occurring in the cosmos.
To begin with, the very concept of time is delineated in classical physics
in terms of correlations between the succession of configurations 
which describe a process and a standard trajectory called the clock. 
The total is an isolated system of fixed energy.
This is relevant for cosmology in that the universe is an isolated
system which we take to be homogeneous and isotropic.
Furthermore, in virtue of the constraint
which arises from reparametrization invariance of time, 
it has total energy zero. Therefore the momentum of the scale factor
is determined from the energy of matter. In the quantum theory
this is exploited through use of the WKB approximation for
the wave function of the scale factor, justified for a large universe.
The formalism then gives rise to matrix elements describing
matter processes. These are shown to take on the form of 
usual time dependent quantum amplitudes wherein the temporal
dependence is given by a background which is once more
fixed by the total energy of matter. 
\vskip 1.5 truecm

\newpage

{\large {\bf Preface}}
\vskip .2 truecm 
\noindent 
This paper is conceived as a pedagogical review to 
the problem of time in quantum cosmology. However,
rather than providing a complete survey of the topic,
it aims to present in a concise manner the line of thought 
developed in refs. \cite{HalH}-\cite{BV}
and \cite{wdwgf}-\cite{future2}.
In particular, by applying 
the usual analysis of statistical fluctuations in large systems,
it confirms that matter evolution is parametrized by a mean time
which is determined by its mean energy.
This is in contrast with the alternative approach\cite{banks, kiefer1, kiefer2}
which is based on the fact that the Planck mass is much larger
than any other mass.

The paper is organized as follows.

\vskip .2 truecm 
\noindent {\bf 1. Introduction.}

\noindent 
We present the problem of time in cosmology and time in general,
and then motivate the procedure brought to bear.

\vskip .1cm
\noindent {\bf 2. Classical cosmology.}
%

\noindent
This section contains the Hamilton-Jacobi formulation of classical cosmology.
In general relativity, the  Hamilton-Jacobi equation is a constraint
which specifies that the total energy in the universe must vanish.
Matter is here represented schematically in the form of 
uniformly distributed ``stars''. 
\vskip .1cm

\noindent {\bf 3. Time. }

\noindent
The concept of time in classical physics is developed. The key point is
the use of the conservation the total energy to correlate trajectories of subsystems.
Application to cosmology is then presented in terms of a model containing 
heavy atoms and radiation.

\vskip0.1cm
\noindent {\bf 4. Quantum cosmology.}

\noindent
The quantized version of the constraint equation (called the Wheeler-DeWitt equation)
 and its solutions as WKB wave functions for the
gravitational part are displayed. The matter degrees of freedom are not yet
coupled among themselves; so this section, like the previous one, is preparatory
to deriving the usefulness of time, a parameter which describes the rates of processes.

\vskip0.1cm
\noindent {\bf 5. Quantum transitions in cosmology.}

\noindent
The perturbation theoretic calculation of transition
probabilities induced by coupling radiation to heavy atoms is carried
out and the Golden Rule derived. It is most interesting here to understand the
vital role of the gravitational wave functions in delivering the
familiar time dependent phases. This Section is the heart of the paper in that
the various threads of formal development are woven together to 
give rise to time as we commonly understand it, a temporal parameter
describing evolution.

\vskip0.1cm
\noindent 
{\bf 6. Concluding remarks.}

\noindent 
This paper attacks the question of time from the point of view
of parametrizing rates of matter processes whereas heretofore
emphasis has been put on the wave function of the universe.
We add a remark on this subject and in particular contrast the
cosmological case with a seemingly similar situation, the Stern-Gerlach experiment.

Our approach suggests a way to look at Mach's 
principle from a quantum point of view. 
We close the paper with  a remark on this point.
%

\newpage

\section{Introduction}

A fundamental problem in physics is posed by the concept of temporality in
cosmology, both in verbal and technical terms. Concerning the former: what is
the clock that times cosmological processes? First of all, what is a clock?

The physicist's answer (and perhaps, in the last analysis, the only one) to this
latter query is: a clock executes a standard classical trajectory and to define
time is to establish a correspondence, or better a correlation, between any
trajectory or any process--the totality of changes in a complicated system--and
points on this standard. A preferred clock is one  whose trajectory allows for
the expression of the laws of physics in simple and intelligible terms. This is
accomplished through the use of perfect periodic systems. These clocks play a
privileged role because they execute a ``wrapped-up" inertial trajectory--a
piecewise recording of this latter.
[One may think of a long ribbon that is
wound up by attaching one of its ends to the needle of an ordinary clock. The other
end is on an inertial trajectory, in the idealization of neglecting the spatial
dimensions of the clock.
In Section 3 more detail will be given on the nature of
periodic systems which will lead to some qualifications
concerning the precise definition of a preferred clock.]
 Because inertial trajectories are simply represented in an 
inertial system, such clocks
are convenient. Indeed Newton's law $F=m \alpha$
relates quantities, 
$F$ and $\alpha$,
 which express first order deviations from
inertiality. Its simplicity and intelligibility are consequences of adopting
what we can well call ``inertial" time--since it issues, as we have said, from
use of a standard trajectory which, at bottom, is inertial.

In this paper cosmology is reduced to its
most essential expression, the study of processes in
homogeneous and isotropic geometries.
We develop the concept of cosmological time in that context
since already at that level of simplification
non trivial conceptual questions arise.
Generalization to situations wherein all gravitational
degrees of freedom are taken into account is very difficult
and has not yet arrived at the same degree of clarity.
For homogeneous geometries, our neat convenient clocks are 
not available as timers. Indeed in the
primitive universe, before the formation of structure, such clocks are nowhere
present. Upon a little thought, one realizes that the only available clock is the
scale factor $a$, or any dynamical function of $a$ which also shows 
semi-classical behavior. 
After all, cosmological data are generally
delivered in terms of the red shift $z$
where $z + 1 = a_{reception} /
a_{emission}$. Cosmology may be summed up as the succession of configurations of
matter which evolve as $a$ does. In its simplest version this is encoded in the
cosmological 
temperature, $T(a)$. The problem is thus: how does one first obtain the
correlations to $a$ and then translate this correlation  into the usual
language of temporal evolution?

The technical aspect of the problem is the one that will deliver the answer. At
first sight one is rather perplexed because in general relativity the problem
of cosmology is reduced to the realization of the single mathematical constraint:
the total energy, matter plus gravity, is strictly equal to zero\cite{mtw}. 
This proposition is a direct
consequence of the fact that the temporal co\"ordinate in general relativity is
not prescribed, there being no preferred frame. (In technical jargon this is
called reparametrization invariance.) Consequently, on the most fundamental of
grounds, one loses the ordinary sense of temporality. Translated into quantum
mechanical terms, the universe is in an energy eigenstate (or perhaps a mixed
state) of zero total energy\cite{WDW}, literally a stationary state\cite{pw}. 
No time!
The equation of constraint imposed on
 this state vector is the Wheeler-DeWitt equation (WDW).

But it is precisely for the same mathematical reason--vanishing total energy--that 
one can recover time since it is this constraint that brings about the
correlations between $a$ and the configurations of matter. In its quantum form,
it is the wave function that encodes these correlations. So it is the job of
the physicist to use the formal theory of constraints, in a perspicacious way, to
convert it into the usual description of temporality. 
It is noteworthy that almost all the efforts on this problem have been 
carried out in the quantum framework. This for the good reason that,
for the quantum 
physicist, it is the fact that the wave function is a stationary state which
gives rise to the acuity of the problem in no uncertain terms.
This is not to say that the problem cannot also be analyzed
in purely classical terms such as in \cite{salopek, barbour, future}
or as in Section 3 of the present paper\footnote{
We are grateful to Julian Barbour for pointing us his work\cite{barbour}
wherein the point of view exposed is similar to that expressed in Section 3.}.

It is the quantum point of view which has been adopted 
in the seminal works of Lapchinski and Rubakov\cite{rub},
of Banks\cite{banks} and of Halliwell and Hawking\cite{HalH}.
To put into evidence time as a measure of matter correlations to $a$,
these authors postulate that the gravitational part of the wave
function of the universe (that part that depends on $a$ alone) is described by a
WKB function. From the action which appears in its phase one makes contact with
classical mechanics in its Hamilton-Jacobi guise. And this, in turn, gives
rise to time $(\equiv t (a))$ as a definition. The usefulness of
this definition is that, by trading correlation to $a$ against correlation to
$t(a)$, one recovers the usual temporal description of the evolution of matter.
Banks calls it a ``maquillage". 

There has been considerable effort on the part of several 
people \cite{Hal}-\cite{ortiz} to implement
the program of refs. \cite{rub, banks, HalH} 
so as to take into account the fact that gravity is driven by matter at
the quantum level as well as to determine how matter evolution is 
modified by this coupling.
The present paper is written in the same vein but whereas previous work
has been concentrated on elucidating ``the wave function of the universe''
we here pursue, following \cite{wdwgf}-\cite{adiab},
a more operational approach; 
to wit: time is a variable which is used to parametrize rates of material processes.
In quantum mechanics, these are given by the ``Golden Rule''
or its non-perturbative generalization. We focus our effort on this aspect of the
problem. How is it that the ``maquillage'' encoded in $t(a)$ yields
the conventional temporal description of material processes ?

In answering this question, certain approximations must be made (in addition
to the idealization of working with homogeneous isotropic universes,
a problem which we do not touch upon, but which clearly requires
further critical analysis\cite{HalH, kiefer2, wdwgf}).  
We show that all these approximations
are justified when the universe contains a large number of particles $N$.
Indeed, by dropping correction terms of order $N^{-1}$
in matrix elements describing matter processes, the metric which appears
is driven by the mean matter energy. And this is precisely what is necessary
to establish the concept of classical time in cosmology.
In this approach the difficult points\cite{vil, isham, wdwin, future2}
of interpreting the wave function 
of the universe are not encountered. 

In summary, the substance of this paper is to show
how the WDW equation encodes changes in matter configurations
as the universe expands.
To this end, we introduce
 interactions among matter degrees of freedom 
and then show, perturbatively, how to recover time
dependent perturbation theory in terms of $t(a)$, wherein 
this latter is determined
from Einstein's equations driven by the mean matter energy.

The main physical principle which we use in 
this analysis
is adiabaticity. By this we mean in its extended
sense, such as the theory of 
non-elastic electronic transitions
induced by molecular motion or 
the quantum description of 
the Stern-Gerlach experiment, 
most elegantly presented by Gottfried\cite{gott}.
In the present case, the validity of the adiabatic scheme
requires that the universe must contain a 
large number of particles.
Indeed, this is used in three
different ways:

\begin{itemize}
\item[a.] To justify the WKB approximation for the gravitational
 wave functions.
\item[b.] To legitimize the conventional expression 
for transition amplitudes. The key point here is that the phase differences
which govern these amplitudes contain differences of macroscopic 
quantities (i.e. $O(N)$) which expanded in power of $1/N$ yield
the usual time dependent phase evaluated 
in the mean background geometry.
\item[c.] In order for this mean to have meaning, it is necessary
that the total matter energies $E$ be dispersed such that
$\langle (\Delta E)^2 \rangle /\langle E \rangle^2 <\!\!< 1$, as is the case in usual statistical 
mechanics. 
In this case, the background geometry is indeed the solution of the 
semi-classical theory of gravity wherein gravity is driven by the mean 
energy $\langle E \rangle$.
\end{itemize}
In this we recover, in a quantum setting, a deep appreciation and a precise
statement of Mach's principle. It is the use of the fact that the universe is 
macroscopic that allows for the
reduction of the interacting gravity-matter system to a description of material
events in a gravitational background, wherein history is given in terms of inertial
time. Specifically, microscopic processes involving a few degrees of freedom 
are so described. Were the universe not large, this would not be possible.
This is as Mach would have it. 
The background, our background, is determined by the myriad of fixed
stars. And the laws of physics, 
expressed in their usual spatio-temporal form, follow accordingly.

\section{ Classical Cosmology}

In this section classical cosmology is presented in a manner that prepares the
way for its quantization. Our treatment
is based on Hamilton-Jacobi theory at
fixed energy.

What follows are the
bare bones necessary to construct the simple cosmological models, Friedman
type universes characterized by Robertson-Walker geometries, homogeneous and
isotropic. These are conceived as an idealization introduced to describe the
universe in the mean, i.e. spatial average.

General relativity is built upon the dynamics of the space-time metric\cite{mtw}. 
The length element in 3+1 dimensions is

\be
 ds^2 = - g_{00} d \xi^2 + 2 g_{0i} d \xi dx^i + g_{ij} dx^i d x^j 
\label{(2.1)}
\ee

\noindent $\xi$ is the temporal co\"ordinate, $x^i$ is spatial with $i = 1, 2,
3.$.

The Einstein-Hilbert action, $S_G= \int  R \sqrt{\vert g\vert } d^3\!x d\xi$ 
wherein $R$ is the scalar curvature, governs the
dynamics of the metric.
One finds that $g_{00}(\xi, x)$ and $ g_{0i}(\xi, x)$ appear without
derivatives with respect to $ \xi$. 
This is analogous to 
electrodynamics 
wherein the time component of the vector potential $A_0$ appears 
without temporal derivatives, essentially
for reasons of gauge invariance. As a result, in the Hamiltonian formalism,
Gauss's law emerges as a condition of constraint. One should not then be
surprised to find similar constraints in the present case. Indeed the momenta
conjugate to $g_{00}$ and $ g_{0i}$ are identically zero, thereby giving rise to
4 local constraints $ {\cal H}_G (\xi, x) = 0, {\cal H}^i_G (\xi, x) = 0$.

When matter is present one must add to $S_G$ its action $( \equiv S_m)$. And,
once again, the absence of temporal derivatives delivers the constraint
conditions, now enlarged to
\ba
 {\cal H}_T (\xi, x)= {\cal H}_G (\xi, x)+ {\cal H}_m  (\xi, x)= 0 &&
\nonumber\\ 
{\cal H}^i_T (\xi, x)= {\cal H}_G^i (\xi, x)+ {\cal H}_m ^i (\xi, x)= 0 &&
\label{(2.2)}
\ea
\noindent 
The quantities $ {\cal H}_m (\xi, x) $ and $ {\cal H}_m ^i (\xi, x)$
are respectively the energy and the momentum densities of
matter. Eq. (\ref{(2.2)}) may thus be interpreted as the vanishing of the sum of the
gravitational and matter energy-momentum densities.
(For the geometric
interpretation of these constraints the reader is referred to \cite{mtw}).
The consequence of this is that gravitational degrees of freedom are
the components of the spatial metric $g_{ij}(\xi, x)$, those which determine the
3-geometry at fixed $\xi$. The space in which these degrees of freedom live is
called superspace. 

The dynamics of homogeneous and isotropic cosmological models 
is carried out in a subspace called mini-superspace.
In this much smaller space, the only part of the $g_{ij}(\xi, x)$ degrees of freedom
which is considered is the scale factor, $a(\xi)$, of cosmology 
which depends on $\xi$ only. It is defined through
the reduced form of the metric
\be
 ds^2 = - N^2 (\xi) d\xi^2 + a^2 (\xi) d \Omega_3 ^2 
\label{(2.3)} 
\ee
 The quantity $ d \Omega_3 ^2$ is the square of the length element on
the unit 3-dimensional hypersurface that is characterized by homogeneous
isotropic geometry. There are only three such: flat, open, closed\cite{mtw}. 
Their differences
will play but a slight r\^ole in what follows. 
In eq. (\ref{(2.3)}), the variable $\xi$ is the
 temporal parameter which parametrizes the different scales realized in the
cosmos. Very important is that $\xi$ may be reparametrized at will, i.e.
$N(\xi)$ is arbitrary. This residual reparametrization invariance is seen by
the explicit expression for the gravitational action when
limited to mini superspace: 
\be
 S_G = {1 \over 2G} \int d\xi N\left[{ a \over N^2}
({d a \over d\xi})^2 + K a 
\right]
\label{(2.4)}
\ee
where $ K = 0, \pm 1$ for the flat, closed and open cases.
$G$ is Newton's gravitational constant. 
This identification comes about by
adding to $S_G$ the action of matter $S_m$. 

To display cosmological dynamics in more general terms, two routes can be
chosen. The first is to fix $N(\xi)$ [said to be gauge fixing] and then solve
for $a(\xi)$ in terms of that particular choice of temporal parameter. The
second method is superior in that it is more physical since it does not resort
to a particular choice of an arbitrary (hence unphysical) temporal parameter.
We shall follow this second method since it will lead us directly to the
concept of time presented in the Introduction. The correlations introduced in
that discussion will be seen to come about from the constraint
which arises precisely because $N(\xi)$ is arbitrary.

The technique used is the Hamilton-Jacobi procedure. One first expresses the
total action $S_T=  S_G + S_m$ in terms of a hamiltonian.
The momentum conjugate to $a$ is 
$p_a  = 
-{ a  \dot a /N G}$
 where $ \dot a = d a / d \xi$. 
For simplicity we take matter to be described by  a uniform
distribution of $n_{star}$ massive stars of mass $M$.
Its action is $S_m = -\int d \xi  N H_m = -\int d \xi  N n_{star} M$.
Then the total action takes the form 
\be
 S_T = \int d \xi [ p_a \dot a - N ( H_G +  H_m)]
\label{(2.5)}
\ee
wherein the gravitational piece is obtained from eq. (\ref{(2.4)}). 
In general there will appear as well, the momenta of matter degrees
of freedom (in the form of $p \dot q$). They are absent
in this simple model 
but the
structure of eq. (\ref{(2.5)}) wherein $ N$ is multiplied by the total hamiltonian
$  H_{T}= H_G +  H_m $ will still hold. 

Explicitly, one obtains
\be
H_G =  {- G^2 p_a^2 - K a^2 \over 2 G a}
\label{(2.7)}
\ee
One sees that $ H_G$ no longer contains $ N(\xi)$ and the same
is true for $  H_m$. This will be
seen when we adopt a specific model later in this section.
Notice the unusual negative sign of  the kinetic energy of $a$.
This is characteristic of theories having reparametrization invariance. 
This negative sign guarantees that there exists always a solution for any
positive matter energy.

Variation with respect to $ N(\xi)$ then yields the residual
constraint that the total energy must vanish
\be
H_T= H_G +  H_m = 0 
\label{(2.6)}
\ee
an equation which delivers $p_a$ as a function of $a$ and the matter
energy. Because of the quadratic character of the kinetic energy of $H_G $,
$p_a$ is determined up to a sign. 
Positive momenta correspond to contracting universes
and negative ones to expanding universes, 
since $p_a = - a \dot a/NG$.

The Hamiltonian-Jacobi formalism is obtained by working within the domain
of possible classical solutions, i.e. which satisfy the constraint. In that
case, it is seen from eq. (\ref{(2.5)}) that the stationary action is 
$S_G = \int^a p(a') d a'$ 
whence $ p(a) = \partial_a S_G $ and the constraint, eq. (\ref{(2.6)}), 
{\it is} the Hamiltonian-Jacobi equation
\be
- G^2 (\partial_a S_G)^2 - K a^2 + 2 G a  H_m = 0 \label{(2.8)}
\ee
This equation is valid for any matter system possessed of a
hamiltonian.

We momentarily postpone further explicitization of the cosmological model of
$H_m$ which we shall adopt to handle quantum transitions in order to first make
contact with the epistemological discussion concerning time which was presented
in the Introduction.

\section{Time}

Whereas, in the cosmological case, energy conservation is derived on the
fundamental grounds of reparametrization invariance, we shall here work in a
more general context wherein energy is conserved (at any value rather than zero
as in gravity) due to the isolation of the dynamical system. Thus,
time dependent forces are absent, so the hamiltonian has no explicit time
dependence.

  We wish to display how it comes about that in such systems one
degree of freedom can serve as a clock, see \cite{arnold}\cite{barbour}. 
It is not amiss here to think of our
earth-sun system idealized to be decoupled from all else in the cosmos. The
clock degree of freedom is the angle of rotation $\phi$
from which 
one extracts time, $t(\phi)$,
 in the conventional sense.
 This is then the temporal parameter
that is used to describe the changing configurations in all the other degrees of
freedom on earth, as has been done since time immemorial. Indeed before
"quartz" time or "cesium" time it was the mean solar time that served as a
standard.

We shall resort to Hamiltonian-Jacobi theory at fixed energy $E$
to display the conversion of correlation of positions into temporality when working
 without external time. 
Before this, we remind the reader 
the way the Hamiltonian-Jacobi function $ S(x,E)$ is used to recover the
 Newtonian temporal
description of a trajectory of a single particle of mass $m$ moving in a 
one-dimensional potential $V(x)$. It is
through the identification of the temporal parameter defined by
\begin{eqnarray}
t (x, E) &=& \partial_E S (x, E) \vert_x = \partial_E
\int^x p (x',E) d x'
= \partial_E \int^x d x' \sqrt{2m (E - V(x'))} \nonumber \\
&=&  \int^x d x' { m \over p(x',E)} = \int^x dx' { 1 \over v (x'; E)} 
 \label{(2.9)}  
\end{eqnarray}
We have introduced the velocity, $v (x, E)$, 
which is here defined by the inverse derivative of the
 $t (x, E)$
with respect to $x$. This is to impress upon the reader 
that this function which expresses temporality is a derivative concept
for isolated systems. This is not how Newton conceived time
which he stated was ``absolute'', but whose absolute character 
was neither used nor exhibited. 
Nevertheless, in classical mechanics it is tacitly taken that 
$t (x, E)$ is a universal parameter, used to
parametrize all trajectories.
Thus all trajectories become correlated one to
the other as time flows. Our aim is to investigate how these correlations
come about using the fundamental principle of dynamics without introducing time.
But first let us return briefly to the concepts at issue.

As such, Eq. (\ref{(2.9)}) is simply the definition of some dynamical quantity. Its
operational sense emerges only in the context of its measurement. And for this
purpose, the trajectory in question has to be compared to some other, which, by
convention may be chosen to be the standard, the clock. For example when I am
in a moving train, in order to answer the question when am I at such and such
location, I consult my watch. Otherwise the question is senseless. Any
operational definition of time involves such a comparison. The fact that the two
trajectories are correlated is a wonderful fact and one must ask why. The
answer given below is that the total energy, (clock + system), of an isolated
system is conserved. This makes very good sense. For example if my watch is
subject to uncontrollable forces from the outside (i.e. time dependent forces)
its needle will move erratically as far as I am concerned and it will no
longer be a faithful clock. Let us now formalize this idea.

Consider two degrees of freedom $x_1$ (system) and $x_2$ (clock)
which comprise the total isolated system of total energy $E$. Their 
action is 
\be
S_{Total} = S_1 (x_1, E_1) + S_2 (x_2, E_2)
\ee
 with $E_1 + E_2 = E$, in the limit where the coupling between them is 
negligible (i.e. infinitesimal). 
Nevertheless  we want to find out why their trajectories are
correlated, i.e. for given initial condition (which henceforth will not be explicitly
written) why a specification of $x_2$ fixes
$x_1$. From these actions one finds two \`a priori  independent temporal
functions 
\ba
 t_1 (x_1, E_1) = \partial_{E_1 } S_1 \vert_{x_1}&&
\nonumber\\
t_2 (x_2, E_2) = \partial_{E_2 } S_2 \vert_{x_2} &&
\label{(2.10')}
\ea
So the question we are asking is why $t_1 (x_1, E_1) = t_2 (x_2, E_2)$
when $E_1 + E_2 = E$ is kept fixed ? 

The answer lies in the principle of least action
that is intrinsic to classical mechanics --how to get trajectories from the
action. Indeed, in this case it gives
\be
 \partial_{\epsilon} S_{Total}\vert_{x_1, x_2, E} =  \partial_{\epsilon}  S_1(x_1, \epsilon)
+  \partial_{\epsilon}  S_2 (x_2, E - \epsilon)
= 0
\label{(2.11)}
\ee
 where $E_1 = \epsilon, E_2 = E - \epsilon$. 
Eq. (\ref{(2.11)}) is the
answer to the question: for what values of the energy split, $\epsilon$, do two
dynamical systems of total energy $E$ pass through the points $x_1$ and $x_2$ at
fixed initial conditions ?

The crucial concept that has been exploited is the notion of a 
{\it single} isolated system, albeit comprised of components parts.
A priori, these parts could be completely independent mechanical 
systems, with no possibility of communications between them 
throughout their history.
Were this so, the conditions we have set down would be operationally
meaningless in that there would be no way of specifying initial conditions
simultaneously. We have tacitly assumed this to be possible, either through
contact or exchange of light rays or what not. Once this is admitted,
then we are dealing with a single system, hence with one independent 
specification
less than that which is required for two completely independent systems.
In short, some interaction among the component parts is assumed, 
be it ever so small. Thus the partitioning of energy $\e$ introduced in eq. (\ref{(2.11)})
is one too many to determine the action of the isolated system.
Therefore, the least action principle requires that the total action
be stationary with respect to its variation, i.e. eq. (\ref{(2.11)}).
This is the dynamical principle which is operationally meaningful,
that one must call upon to replace the Newtonian sense of absolute time.

In this fashion one sees that the description of the trajectory of the system
(1) is given by the function $x_1 (x_2)$. Moreover, it is delivered through eq. 
(\ref{(2.11)}) in the
form $x_1 (t_2 (x_2))$. The usefulness of $ t_2 (x_2, E_2)$ rather that $x_2$,
itself, is that it opens the way to express the dynamics of $x_1$ without
appeal to the clock's dynamics. For periodic systems with one degree of freedom,
this is accomplished by adopting, for this latter,
 action-angle variables $( \equiv J_2, \theta_2)$
since $\theta_2$ is directly proportional to $ \partial S_2/ \partial E_2$
because the angular velocity ($= (\partial_E J)^{-1}$) is independent
of $\theta_2$. Hence the
previously displayed correlation of $x_1$ to $x_2$
delivers the trajectory directly in terms of
$\theta_2$ instead of some complicated function $t_2(x_2)$. Were there many
clocks, each described by an angle $\theta_i$, then the stationarity condition
correlates them so that any angle,
 $\theta_i$, can be used to parametrize the
trajectory of $x_1$. Indeed this is a universal function given by $ x_1
(\theta_i \partial_{E_i} J_i)$. It is $ (\theta_i \partial_{E_i} J_i)$ 
that is called time\footnote{
For periodic systems $(2)$ of more than one degree of freedom,
one must replace the above consideration by more complicated expressions
which take into account the shape of the trajectory, 
for example the shape of the
earth's trajectory around the sun. Indeed the angle traversed by the 
earth with respect to fixed stars does not define ``time''.
However $n$ complete revolutions of the earth is a 
faithful measure of time.}. Its universal
character is now demonstrated. 
For simplicity, the above has been presented for the case where  system 
$(1)$ has one degree of freedom. Generalization to systems
possessing many degrees of freedom is straightforward.

We now apply this approach to matter evolution in a cosmological
situation. For purposes of
describing typical quantum transitions in cosmology, we take matter to be
composed of conformal radiation and heavy particles. These are explained in turn.

Conformal radiation refers to massless fields. This is exemplified in nature by
photons and we shall take a scalar version of them. Conformality has as
consequence that the photon's energy, as measured in ``conformal'' time $\eta$, 
see below for its definition,
is proportional to its inverse wavelength
\be
\omega_{ k} = \sqrt{\sum_i  k^2_i}
\label{(2.14)}
\ee
For simplicity we have taken the case of flat 3-geometries
so that 
the components
$k_i$ label the non dimensional wave numbers. Thus $
\omega_{ k}^2$ is the
eigenvalue of the 3-laplacian, $\bigtriangleup_3$,
 non dimensionalized by the scale factor $a$, 
corresponding to the metric of eq. (\ref{(2.3)}).
The definition of conformal time is obtained from this equation with the choice
$N(\eta) = a(\eta)$ so that the 
d'Alembertian operator is $\partial_\eta^2 - \bigtriangleup_3$.
The solutions for the flat case are thus $exp\ {i( \om_k \eta -{\bf k}\cdot
{\bf r})}$ 
with the dispersion relation eq. (\ref{(2.14)}) where $r =({\bf r \cdot \bf r})^{1/2}$
is the non dimensional length obtained from $d\Omega_3$ of eq. (\ref{(2.3)}).

We now turn to the description of ``heavy'' particles.
By heavy we mean that they are sufficiently massive so as not to be produced by
the cosmological expansion\cite{BD}. 
For this to be true it is required that
\be
 M \gg ( { d a/ dt \over  a}) 
\label{(2.12)} 
\ee
where $\hbar = c = 1$ and 
 $t$ is the proper time corresponding to the choice $N = 1$ in eq. (\ref{(2.3)}).
The energy relative to $t$ of
a heavy particle whose conserved wave number components are $p_i$ is given by
\be
 \Omega (t) = \sqrt{M^2 + \sum_i p_i^2/a(t)^2} 
\label{(2.13)}
\ee
where the dependence on $a$ is understood in the light of eq. (\ref{(2.14)}) 
valid for massless particles
when it is noticed that for that  case, $\Omega = \omega / a$.
 It is the time
dependence of $\Omega $ through $a(t)$ which 
generates pair production in that $M$ and
$p/a$ ``feel" the expansion differently. 
For simplicity in this and the next
section we shall set $p = 0$.

In resum\'e the action of matter which is the subject of 
Sections 4 and 5 is
\be
S_{matter} = S_M + S_\gamma = {1 \over 2} \int d \xi N
\{ [ { \dot \psi^2 \over N^2 }- M^2
\psi^2] + a^{-1} [{ a^2 \dot \phi^2 \over N^2 }- \omega_k^2 \phi^2 ] \} 
\label{(2.15)}
\ee
Dots mean $d/d\xi$. As compared to conventional notation, we have
simplified by rescaling the matter field $ \psi$ by a factor of $a^{3/2}$
(hence $ \psi$ has dimension $ L^{1/2}$) and the radiation field $ \phi$ by a
factor of $a$ (hence $ \phi$ is dimensionless). The scaling of $\phi$ results in the simple form, 
eq. (\ref{(2.15)}), in consequence of conformal symmetry.
Instead that of $\psi$
introduces terms in $\dot a$ which lead to pair production, see \cite{BD, adiab}.
In this paper, due to eq. (\ref{(2.12)}), they have been 
neglected.
For simplicity we work here with a single value of $\omega_k$ and introduce its
continuum spectrum in subsequent sections to display transitions in usual terms.

With these simple choices, $ \psi(\phi)$ are harmonic oscillators in $t (\eta)$
respectively. Denoting their constant amplitude by $ A_M (A_\gamma)$, their energy (with
respect to $t$) is
\ba
 E_m (a) &=& M \vert A_M \vert^2 + \omega_k  \vert A_\gamma \vert^2 /a
\nonumber\\
&=& \e_M + \e_\gamma /a
\label{(2.16)}
\ea
Since both $\e_M $ and $\e_\gamma $ are conserved, i.e. independent of $a$,
the total action is a sum of three actions
\be
S_T( a, \psi, \phi) = S_G( a; \e_M , \e_\gamma) + S_M (\psi, \e_M )
+ S_\gamma (\phi,  \e_\gamma) 
\label{sum3}
\ee
wherein the gravitational part
satisfies the Hamilton-Jacobi equation, 
eq. (\ref{(2.8)}), with $H_m =  E_m (a)$.
The solution is
\be
 S_G (a; \e_M , \e_\gamma) = \int^a da' \ p (a'; \e_M , \e_\gamma) \label{(2.18)}
\ee
 where $p(a; \e_M , \e_\gamma)$ is the momentum of gravity driven by the matter energy
specified by $\e_M$ and $\e_\gamma$:
\be
 p (a; \e_M , \e_\gamma ) = - G^{-1}  [ - K a^2 + 2aG (\e_M + \e_\gamma /a) ]^{1/2} 
\label{(2.19)}
\ee
The sign is chosen to correspond to $\dot a > 0$, an expanding
universe.

To obtain cosmological time we refer to eq. (\ref{(2.11)})
but generalized to include matter in two forms, $\psi$ and $\phi$
coupled to gravity through their energy, eq. (\ref{(2.16)}).
Because the matter action is a sum of two terms it is now necessary
to consider extremization with respect to the independent constants
$\e_M $ and $\e_\gamma$. Variation with respect to the former correlates 
$\psi$ to $a$. As emphasized before, this correlation between dynamical 
variables is delivered in terms of the temporal parameter 
conjugate to $\e_M $, hence given by
\ba
t(a; \e_M , \e_\gamma) &=& -\partial_{\e_M } S_G (a; \e_M, \e_\gamma)
= -\int^a da' {  a' \over G p (a'; \e_M, \e_\gamma)}
\label{tdef}
\ea
Similarly variation with respect to $\e_\gamma$ correlates $\phi$ to
$a$ through $\eta(a; \e_M , \e_\gamma)$ given by
\ba
\eta
(a; \e_M , \e_\gamma) &=& -\partial_{\e_\gamma}S_G (a; \e_M, \e_\gamma)
= - \int^a da' {  1 \over G p (a'; \e_M, \e_\gamma)}
\label{etadef}
\ea

To conclude, we have made explicit how time arises through correlation
of the motion of different degrees of freedom within an isolated
system. Among these there is one which can be used to define the time. 
This is called the clock. Periodic clocks are most useful, but they are not always
available such as in cosmology where the only natural clock is $a$.
The price to pay is that it entails a quadrature which encodes
the unfolding of the dynamics up to the present.
Indeed, this is one of the central points of debate in realistic
cosmology. 
What is the function $p (a; \e_M , \e_\gamma)$ 
and hence what is the age of the universe?

In the next Sections we shall show how in quantum developments
the classical times introduced in eqs. (\ref{tdef}, \ref{etadef}) emerge
from the formalism quite naturally once transitions
are induced by interactions\footnote{
This raises the question whether one can implement the same program 
purely classically. How does one find the history of the effects of interactions
in classical terms ? This has been carried out in \cite{future}.}.

\section{Quantum Cosmology}

The classical constraint, eq. (\ref{(2.8)}),
 can be quantized in the usual way, up to an
operator order ambiguity often encountered in passing from classical to quantum
theory. This ambiguity occurs in $H_G$ and we shall adopt the simple rule of
setting $p_a^2$ to $ \partial^2_a$. There is some polemic in the
literature on this point which is tied up with the interpretation of the ``wave
function of the universe"\cite{vil, isham, wdwin, future2}. This 
review is not concerned with this profound aspect
of quantum cosmology, but rather with the behavior of matter in 
the universe\cite{wdwgf, kiefer2}. We
postulate that the gravitational wave functions are WKB and 
justify this \`a posteriori by showing that corrections are negligible in a
universe containing many particles. Since WKB wave functions are insensitive
to the ordering problem we need go no further into the question. We
may remark that on this basis questions concerning interpretation and ordering
seem more to do with the very beginning of the universe, rather than that which
is covered by the part of cosmological history which is, at least partially,
accessible to experimental and theoretical analysis. And of course one must
bear in mind that in any quantum theory there are always the deeper
epistemological problems which, at least at present, seem to have nothing to do
with cosmology. These will not be touched upon in this either. As Gottfried 
says\cite{gott}
we operate as ``quantum mechanics".

The quantum version of eq. (\ref{(2.8)}), called the 
Wheeler-DeWitt (WDW) equation, is thus
\be
 [H_G + H_m]\;\Xi (a, \psi, \phi) = [ G^2 \partial_a^2 - K a^2 + 2 G a H_m]\;
\Xi (a, \psi, \phi) = 0 \label{(3.2)}
\ee
$\psi$ and $ \phi$ are oscillator co\"ordinates appearing 
in the matter hamiltonian $H_m$, see eqs. (\ref{(2.15)}, \ref{(2.16)}).
Matter is quantized in the usual way. The  eigenstates of 
our simple model for $H_m$ are 
products of states of two harmonic oscillators. Thus, the eigenvalues are
\ba
  E (a; n_M, n_\gamma) &=& \expect{n_M \vert \langle n_\gamma }{H_m(a)}{ 
n_\gamma \rangle \vert n_M} 
\nonumber\\ &=& (n_M + 1/2) M + (n_k + 1/2) (\omega_k/a)
\label{(3.1)}
\ea
Since these eigenstates are stationary, i.e. $\partial_a \ket{n} = 0$,
the general solution of eq. (\ref{(3.2)}) 
can be written in the form
\be
 \Xi ( a, \psi, \phi) = \sum_{n_M, n_\gamma} \; c_{n_M, n_\gamma} \; \Psi
(a, n_M, n_\gamma) \bk{\psi \vert n_M} \bk{\phi \vert n_\gamma} \label{(3.3)} 
\ee
 where the gravitational wave function $\Psi(a, n_M, n_\gamma)$ 
corresponding to the matter state 
$\ket{n_\gamma}\ket{n_M}$ obeys
\be
 [ G^2 \partial_a^2 -  Ka^2 + 2 G a E(a; n_M, n_\gamma)] \Psi (a, n_M, n_\gamma)
= 0 \label{(3.4)}
\ee 
It is a second order equation of an operator having a
continuous spectrum, so the amplitude of the solutions
is determined by the Wronskian (W) (often called the
current, here the expansion of the universe). The sign of W is chosen to
correspond to an expanding universe and its absolute value fixed to be unity.
At this stage, this is convention, however upon introducing interactions
among the matter states $\ket{n_\gamma} \ket{n_M}$, it 
will be seen that the Wronskians generated by $\Psi (a, n_M, n_\gamma)$
determine 
$a$-dependent probability amplitudes which give rise to the usual
statistical interpretation of quantum mechanics subject to the condition\cite{wdwgf}
\be
W = \Psi^* (a; n_M, n_\gamma)\; i {\stackrel{\leftrightarrow}{\partial_a}} \;
\Psi (a; n_M, n_\gamma) = 1 \label{(3.5)}
\ee

The WKB solution of eq. (\ref{(3.4)}) with unit Wronskian is
\be
 \Psi (a; n_M, n_\gamma) = {1 \over \sqrt{2 p(a; n_M, n_\gamma)}} \; 
\exp \left[ i \int_{a_0}^a \; p
(a' ; n_M, n_\gamma) d a' \right]
\label{(3.6)} 
\ee
for all $n_M, n_\gamma$.

The value $a = a_0$ is an arbitrary
reference point. The value of the classical momentum
 $p(a, n_M, n_\gamma)$ is read off from eq.
 (\ref{(2.19)}) with $E_m(a) = \e_M + \e_\gamma /a $ replaced 
its quantum expression given in eq. (\ref{(3.1)}).
 We do not enter here into the question of how to handle closed
universes near turning points $(p(a; n_M, n_\gamma) = 0$ when $ K = 1)$. 
The material
presented here is preparatory to the description of transitions in matter
immersed in an expanding universe (hence the expanding phase of a closed
universe). So  what we have in hand is sufficient provided $n_M$ and $n_\gamma$
are large enough to keep the universe expanding at the value of $a$ in question.
We postpone the discussion of the legitimacy of the WKB approximation in
eq. (\ref{(3.6)}) to the end of next
section.

Let us now briefly show
how usual quantum mechanical ideas
and formulae emerge in this scheme. 
Instead of using the configuration
representatives $ \bk{\psi \vert n_M }$ and $\bk{\phi \vert n_\gamma }$ 
as in eq. (\ref{(3.3)}), it will
be convenient to work with $\ket{ \Xi }$ as a state (a ket) in Fock space:
$\ket{\Xi (a)} =
\sum_{ n_M, n_\gamma}
c_{ n_M, n_\gamma} \Psi(a; n_M, n_\gamma) \vert n_M \rangle \vert n_\gamma  \rangle $.
 From eq. (\ref{(3.5)}),  we find
\be
c_{ n_M, n_\gamma} =  
\langle n_M \vert \langle n_\gamma \vert \Psi^*(a; n_M, n_\gamma)
  i {\stackrel{\leftrightarrow}{\partial_a}}
\vert \Xi (a) \rangle
\label{(3.8)}
\ee
and one checks $ \partial_a c_{ n_M, n_\gamma}  = 0$. 
This of course is a consequence
of our model with no matter interactions so that $n_M, n_\gamma$ 
are constants
independent of $a$. Thus the normalization condition $ \sum_{ n_M, n_\gamma}
 \vert c_{ n_M, n_\gamma}  \vert^2
= 1$, being conserved, can and will be adopted. 
In order for matrix elements to take on the familiar form 
consistent with the statistical interpretation of 
quantum  mechanics, we make use of eqs. (\ref{(3.3)}) and (\ref{(3.5)}).
For example, the mean value of the energy is given by
\be
\bk{H_m(a)}_{\Xi} = 
\langle  \Xi (a) \vert \left[
- i {\stackrel{\leftarrow}{\partial_a}}  H_m(a) 
+  i H_m(a) {\stackrel{\rightarrow}{\partial_a}} \right]
\vert \Xi (a) \rangle = \sum_{ n_M, n_\gamma}  \vert c_{ n_M, n_\gamma}  \vert^2 \;
E(a; n_M, n_\gamma) \label{(3.9)}
\ee 
In this way,
the $c_{ n_M, n_\gamma}$
are the conventional probability amplitudes to find the matter in state 
$\vert n_M  \rangle\vert n_\gamma \rangle$, in
the universe. With our choice of Wronskians, gravity does not enter into
these diagonal elements, save for the $a$ dependence 
of the eigenvalues $E(a; n_M, n_\gamma)$. This is in analogy to 
$e^{-iEt}$ dropping out in quantum mechanics. 
That we can construct this conventional result from the solutions
of the WDW equation is to be expected. Indeed, when applying the 
same procedure in more conventional quantum settings such as
atomic physics in a Stern Gerlach experiment\cite{gott}, 
or the transitions of accelerated systems\cite{rec} giving rise to the
Unruh effect, the same types
of expressions follow. 

\section{Quantum Transitions in Cosmology.}

Our aim is to apply the formalism 
of the previous Section to recover the
time dependent perturbative theory of transition rates --the Golden Rule.


In quantum cosmology however, since there is no time, there is no Schr\"odinger 
equation which delivers the sense of evolution of amplitudes.
This physical idea of propagation must somehow or other be encoded
in the constraint in its quantum version, the WDW equation.
In Section 4, we have seen that, in the absence of interaction,
the formalism was set up to express the physical idea that, the
$c_{n_M, n_\gamma}$, the 
probability amplitudes to find a certain number of matter quanta
were fixed, i.e. independent of $a$.
We now ask the following question. Suppose we know these
amplitudes at a certain value of $a$ ($=a_0$).
From this datum, how does the constraint equation deliver the values of the
amplitudes at other values of $a$? If the hamiltonian interaction $H_{int}$
depends on $a$,
the linear combination which represents the complete vector
state $\ket{\Xi(a_0)}$ in Fock space of free matter states at $a_0$ 
will contain different amplitudes in this space at different values of $a$.
We seek the law which governs the variation of these amplitudes,
i.e. the description of evolution, a succession of 
events which are quantum transitions.

As shown in \cite{wdwgf, wdwin}, the manipulation of the WDW equation in the 
presence of $H_{int}$ leads to expressions which resemble conventional
time dependent theory. Indeed, the momentum of gravity becomes
operator valued in Fock space --in total analogy to the interaction representation
of the evolution operator in usual time dependent perturbation theory.   
Then in the integrals over $a$ that come up
 (which are in fact the usual integrals
over time necessary to pick up a resonance) in transition amplitudes,
the difference of WKB phases of the gravitational
wave functions associated with the initial and final matter states delivers what 
is usually called the virtual energy. 
Necessarily, since it
is the energy constraint that is sacrosanct in all of this.
In this way, it will be seen that the use of a classical background in usual time
dependent theory is replaced by the 
recoil effects of gravity necessitated by the energy constraint.

The model used in this section to describe matter is a 2-level atom of masses
$m$ and $M$ where $(M-m)/ M = \Delta m/M \ll 1$ and positive. The atom
interacts with conformal radiation (scalar photons). Thus
\begin{eqnarray}
\tilde H_m &=& H_m + H_{int} \nonumber \\
H_m &=& M d_M^\dagger d_M + m d_m^\dagger d_m + 
\sum_k (\omega_k/a)  d_k^\dagger
d_k\nonumber \\
H_{int} &=& g \sum_k \  \psi_M \psi_m  \;\phi_k 
\label{inter}
\end{eqnarray}
where the two atom states have been described 
by two massive fields and where
$\phi_k$ is the $k^{\underline{th}}$ component
of the photon field, see eq. (\ref{(2.15)}).
The $d_i$ ($d_i^\dagger$)
are annihilation and creation operators of the three fields, {\it c.f.} eq. 
(\ref{(3.1)}). 
The momentum of the atom is neglected, as is often done in
atomic physics for heavy atoms wherein $\Delta m \ll M$.

We remind the reader how to recover the Golden Rule in lowest order
perturbation theory. In a
fixed cosmological background described by $a(t)$, the Schr\"odinger 
equation for the
expansion coefficients, $c_n$, of the wave function in the basis of the
eigenfunctions of $H_m$ is
\be
 i {d c_n \over dt} = \sum_{n'}  c_{n'} \expect{n'}{
e^{i \int^t_{t_i} H_m (t^\prime) dt^\prime} \;  H_{int} \;
e^{-i \int^t_{t_i} H_m (t^\prime) dt^\prime}}{n}   
\label{(4.2)}
\ee
where $n$ and $n'$ designate triplets of occupation numbers
$n_m, n_M, n_k$ characterizing uninteracting matter states.
Solving eq. (\ref{(4.2)}) to lowest order
in $g$ subject to $ c_n (t_i) = \delta_{ni}$ gives the transition amplitude
\be
{\cal A}_{fi} = \int_{t_i} ^{t_f} dt^\prime \expect{n_f}{H_{int}}{n_i} 
\; e^{ i \int^t_{t_i} [ E_f
(t^\prime) - E_i (t^\prime) ] dt^\prime }
\label{(4.4)}
\ee
For example, the amplitude of emission of photon $k$ is proportional to
\be
 i g \sqrt{n_k + 1} \int_{t_i} ^{t_f} dt \;
e^{ i [ \Delta m (t - t_i) - \om_k (\eta (t) - \eta (t_i) ] }
\label{(4.5)}
\ee
where the phase referring to the photon is $\int^t_{t_i} dt^\prime /a(t^\prime)
= \eta (t)-\eta (t_i)  $.
Conventional Golden Rule physics is recovered if $a(t)$ varies sufficiently
slowly during the time necessary to pick up resonant conditions, i.e. $(\dot
a/a) \ll \Delta m$. Then one can set $a = a(t_i)$ and the usual technique
of extracting a rate applies. One squares and integrals over $k$ with the
correct phase space factor. The resonant values of $\om_k$ dominate 
$\om_k / a(t_i) =
\Delta m$ corresponding to energy conservation around $t_i$ and the
Golden Rule rate formula ensues upon integrating over $k$.

Our aim now is to show how the formalism of Section 4 recovers 
eq. (\ref{(4.2)}). The rest
is conventional quantum mechanics as given in the above example. To this end one
introduced the following trick. Rather than working with the expansion
coefficients $c_n$ of eq. (\ref{(3.3)}), which now will depend on $a$ owing to $H_{int}
\neq 0$, we introduce new coefficients $ \tilde c_n$ defined by
\be
 \tilde c_n (a) \equiv  \bra{n} \Psi^* (a, n ) \; i
{\stackrel{\leftrightarrow}{\partial_a}} \ket{ \Xi (a)} \label{(4.6)}
\ee
One verifies that $ \tilde c_n$ and $c_n$ differ by a term in $dc_n/da$
hence by terms in $O(H_{int})$. We shall estimate this difference subsequently
and show that it is negligible for a macroscopic universe. 
%
The advantage of working with $\tilde c_n$ is the simplicity of its equation of
motion which then leads directly to eq. (\ref{(4.2)}). 
Indeed, one has
\be
-i \partial_a \tilde c_n = \bra{n} \partial_a^2 \Psi^*(a, n ) - \Psi ^*(a, n )
\partial_a^2 \ket{\Xi (a)} \label{(4.7)}
\ee
We now substitute  eq. (\ref{(3.4)})
for $ \partial_a^2 \Psi ^*$
 and eq. (\ref{(3.2)}) for $\partial_a^2  \Xi (a)$
where $H_m$ in that equation is here replaced by $\tilde H_m$
of eq. (\ref{inter}). 
Whereas the former does not contain $H_{int}$, the
latter does, since the constraint equation (WDW) contains the \underline{full}
matter hamiltonian. Thus the momentum $p_a$ carried by $\ket{\Xi (a)} (= i
\partial_a \ket{\Xi (a)}$) is operator valued in Fock space. It contains the
non diagonal matrix, $H_{int}$. Gravity is thus participating in a highly
non-trivial way in matter dynamics through the constraint condition.
Substituting  eq. (\ref{(3.4)}) and eq. (\ref{(3.2)}) into eq. (\ref{(4.7)}),
one notices that the term in $-Ka^2$
drops out. Similarly were a cosmological constant present it would cancel as
well. 
The result is
\begin{eqnarray}
i \partial_a \tilde c_n &=& {2a \over G} \sum_{n'} c_{n'} \expect{n'}{H_{int}(t)}{n} \
 \Psi^* (a, n) \Psi (a, n')  \nonumber \\
&\simeq& {2a \over G} \sum_{n'} \;\tilde c _{n'} \expect{n'}{H_{int}(t)}{n} \
\Psi^* (a, n) \Psi (a, n')
\label{(4.8)}
\end{eqnarray}
In the second equality of eq. (\ref{(4.8)})
 we have written $ \tilde c_{n'}$ for
$c_{n'}$ in anticipation that the difference is small. 

Eq. (\ref{(4.8)}) is a nice equation in that $ \sum \vert {\tilde c}_n (a) \vert^2 = 1$ is
conserved owing to the hermiticity of the matrix $\expect{n'}{H_{int}(t)}{n} 
\Psi^* (a, n) \Psi (a, n')$. 
And since $ \tilde c_{n}(a)$ is to very good approximation equal
to $c_{n}(a)$ we have unitary evolution. 

Adopting the WKB approximation for $ \Psi (a, n)$ we then have
\be
i  \partial_a c_n =  \sum_{n'} \; c _{n'} 
\expect{n'}{H_{int}(a)}{n}  {a  \over G
\sqrt{p(a; n) p(a, {n'})}} \; \exp \left[-i \int^a_{a_i} [p (a^\prime; n' ) - p(a^\prime; n)] 
da^\prime \right] 
\label{(4.9)}
\ee
In an objective sense, eq. (\ref{(4.9)}) is the central result
of this paper: It describes the unitary evolution of material processes 
as a function of $a$. 
 Its derivation
turns around the Wheeler-DeWitt energy constraint.
This is used in two different ways: 1) the appearance
of $H_{int}$ which has arisen from the constraint for $\ket{\Xi (a)}$,  2) the
function $\Psi (a, n)$ which encodes unperturbed propagation. The oscillations in
the integral on the phase of eq. (\ref{(4.9)}) are those of the gravitational wave
functions which occur during the time is takes to sort them out to get resonances
$O(1/\Delta m)$. 

This derivation makes clear the essential physics of the nature of
temporality in cosmology. The phase $[ = S_G =
\int p_a da]$ contains the macroscopic
quantity $p(a; n_M, n_m, n_\gamma)$ of eq. (\ref{(2.19)}) enlarged to include the
contribution of the second massive field of mass $m$. For a flat 
universe it is proportional to $\sqrt{E_{tot}}$,
$E_{tot}$ being the energy of all the matter in the universe. This sounds
ridiculous, but the essential point is that it is the difference $ \Delta S_G = \int 
 \Delta p_a da$ associated with the initial and final matter states which
figures in the description of the physical process and this latter is
microscopic in character, albeit taking place in a macroscopic universe.
Similarly $d a$ is the differential of a quantity of cosmological proportions.
But the particular function of $a$ which multiples $\Delta E$, is the microscopic
quantity $\partial_E p_a da = dt $, given by eq. (\ref{tdef}), 
simply the differential of proper time, the temporal
parameter which is intrinsic to the description of the rate process studied in
the rest frame of our comoving heavy particles.

In a subjective historical sense, it nevertheless
is important to make contact with conventional time dependent 
Schr\"odinger theory. Precisely, what are the conditions to
realize 
the replacement of $a$ by $\bar t(a)$
where $\bar t(a)$ is the mean time, i.e. the solution of eq. (\ref{tdef})
evaluated with the mean energy $E(\bar n, a)$.
Unlike the previous considerations which are of purely dynamical
character, these new conditions involve the specification of the
initial data: the set of $c_n(a = a_0)$. Indeed, 
when the fluctuations in $n$
around the mean values $\bar n$ 
are small,
one may expand 
both the prefactors and the phases of eq. (\ref{(4.9)}) to first order 
in the changes $n -\bar  n$ and $n'-\bar  n$, or 
equivalently in $E(n, a) - E(\bar n, a)$. (For 
a discussion of higher order terms, see \cite{wdwgf}.)
Upon the further use of eq. (\ref{tdef})
to express $a da / G p(a; \bar n)$ as $d\bar t$, we obtain 
\be
i ({ G p(a; \bar n ) \over a }) \partial_a  c_n = 
 i \partial_{\bar t} c_n = \sum_{n'} c_{n'}
\expect{n'}{H_{int}}{n}  \; e^{ - i \int_{t_0}^{\bar t(a)}
 [E(a(\bar t'), n') - E(a(\bar t'), n)] d\bar t'} 
 \label{(4.10)}
\ee
It is to be noted that the momentum prefactor ($= p(a; \bar n)^{-1} $
to order $1/N$) is precisely what is necessary to convert 
$\partial_a $ to $\partial_{\bar t}$. Its 
origin stems from the
second order character of the ``free'' WDW equation, eq. (\ref{(3.4)}),
and our choice of the Wronskians, eq. (\ref{(3.5)}).

Therefore using the restriction on initial data that the fluctuations
of $n_M, n_m$ and $n_\gamma$
around their mean values are small, we can replace the 
$n$-dependent functions of eq. (\ref{(4.9)}),
by the conventional expressions governed by 
the mean time $\bar t(a)$.
In this way we have recovered 
the Schr\"odinger equation, eq. (\ref{(4.2)}),
governed by a single time
determined by the evolution of the mean gravitational background.
In conclusion we emphasize that this result so familiar from fixed background
physics was entirely encoded in the product of the two gravitational wave functions 
 $\Psi^*(n_f, a) \Psi(n_i, a)$, see eq. (\ref{(4.8)}). 
This is what delivers the space-time history of events.

It is appropriate here to clarify a confusing point that exists in the
literature. This is concerned with the specification of the gravitational 
background that is
used to extract time from the WKB wave functions. Based on Bank's 
work\cite{banks}, many authors make an expansion 
in the inverse square Planck mass. In the
absence of a cosmological constant, the background so obtained is that of an
empty universe, i.e. characterized by $n_M = n_\gamma =0$.
But from eqs. (\ref{(4.9)}, \ref{(4.10)}) one sees that this is incorrect:
In the microscopic transition $n \to n'$, it is the
background driven by the total matter energy $E(n, a)$ (or $E(\bar n, a)$, 
they differ by $O(\sqrt{N})$ out of $O(N)$) which determines the 
time dependent phases. 
The value of the Planck mass is completely irrelevant.

It remains to estimate the difference between $c_n$ and $\tilde c_n$. From 
eq. (\ref{(3.3)}) and eq. (\ref{(4.6)}), one has
\be
 \vert {\tilde c_n - c_n \over c_n}\vert  = \vert (\partial_{a} \ln c_n) /
p (a; n)
\vert \cong  ( \partial_{a } \ln \tilde c_n) / p (a; n)
\label{(4.12)}
\ee
where in the second equality we have replaced $c_n$ by
$\tilde c_n$ in anticipation that the difference is small. From eq. (\ref{(4.8)}),
the order of magnitude of the
r.h.s of eq. (\ref{(4.12)}) is 
$O( { g \over M } { 1 \over N})$ 
where $g$ is the coupling constant. Since our whole perturbative
scheme depends on the smallness of $g$, one sees that the correction is 
smaller than $1/N$.

It still remains to justify the use of the WKB approximation. 
As emphasized above, the physics of processes in the universe
makes use of the functions $\Psi(a,n)$,  eq. (\ref{(3.6)}), through  eq. (\ref{(4.8)}), 
i.e. in the form $\Psi^*(a,n)\Psi(a,n')$ where $n'- n$ is microscopic in character. 
Therefore it suffices to estimate the error induced by WKB in this quantity. 
After some algebra\cite{wdwgf},
one finds that the local\footnote{When abandoning the WKB approximation, 
there is also a ``global'' effect namely the 
production of back-scattered wave functions describing contracting universes,
see \cite{ce}. 
In\cite{adiab}, it has been proven  
that eq. (\ref{(4.9)}) is correct upon neglecting precisely the contribution of these
back-scattered waves. The physical consequences of backscattering
transitions will be presented in \cite{future2}.}
relative corrections scale like ${1 /N^4}$. 
Thus these corrections are also 
completely negligible in a macroscopic universe.

It is also worth mentioning the close analogy between the preceding
development and the derivation of the canonical distribution from
a microcanonical ensemble in statistical mechanics.
In the latter one considers the distribution of $\e$, the energy of a small system 
immersed in a large one (the reservoir) of energy $E_{tot} - \e$.
Its determination follows the same procedure as the determination
of the lapse of time in a universe as it comes in the passage of eq. (\ref{(4.9)})
to eq. (\ref{(4.10)}), to wit: a first development in $\Delta \e$,
the energy fluctuation of the small system.
In both cases the big system is a reservoir and first order expansions
in the small energy fluctuations yield the parameter that characterizes on one hand 
evolution, $t = \partial_E {\rm Action_{gravity}}
\vert_{E = \langle{E_m}\rangle}$, 
and on the other the distribution,
$\beta = \partial_E {\rm Entropy_{reservoir}}\vert_{E_{total} - \bk{\e}}$.
More detail is given in Appendix B of \cite{wdwgf}.


\section{Concluding Remarks}

In this essay we have shown how the evolution of matter in the cosmos 
is parametrized by the cosmological time $\bar t(a)$ defined
by the expansion of the universe, see eq. (\ref{tdef}).
Events in the cosmos
are so analyzed and indeed this is the only operational 
sense that one can ascribed to the word ``time''.

The outcome of our analysis is the unitary evolution 
described by eq. (\ref{(4.8)}) wherein the expanding universe
is encoded in the WKB approximation to $\Psi(a;n)$.
The all important point is the dependence of $\Psi(a;n)$ on the 
state of matter, the index $n$, which is determined by the
constraint eq. (\ref{(3.4)}). For a macroscopic universe further 
reduction is possible since the microscopic transition
described by $\bra{n'} H_{int} \ket{n}$ engenders but infinitesimal
differences in the WKB phases displayed on the r.h.s. of eq. (\ref{(4.9)}).
To first order in these differences, one finds an equation
which generalizes eq. (\ref{(4.10)}) in that the 
mean time $\bar t(a)$ is replaced by a ``state
dependent time'' $t(a,n)$ defined by eq. (\ref{tdef}) 
with $E_{matter}=E_n(a)$.
Only for a macroscopic universe whose fluctuations 
in energy density are typically statistical (i.e. $O(N^{-1/2})$)
can one replace $t(a,n)$ by $\bar t(a)$, for then $t(a,n)$,
being a microscopic variable, also has relative fluctuation
$O(N^{-1/2})$. On the contrary, when the energy density
is widely distributed, i.e. far from equilibrium, such
a reduction is not possible and one must fall back on the 
initial formulation parametrized by $a$.
 
It is appropriate to compare this operational approach 
with that of \cite{BV, ce} which focussed
on the wave function $\Xi(a, \phi, \psi)$
 rather than on processes encoded in it.
In these works, interactions among matter degrees of freedom 
were not considered; hence  
these works are preparatory in the sense of Section 4.
Thus, as in eq. (\ref{(3.4)}), each matter sector, $n$, defines
its corresponding gravitational wave function.
When these sectors are crowded around
energy eigenfunctions of nearly the same energy,
one may expand the gravitational wave functions
around the mean value,
so as to express the dependence of $\Xi$ in $a$
in terms of the mean time $\bar t(a)$.
However, in the absence of interactions,
this procedure of making use of the quantum spread
 is physically empty since the sectors
will never interfere.

We may note the difference between the above consideration
and the analysis of the Stern Gerlach experiment\cite{gott}.
In that case the center of mass motion is encoded in the
construction of a wave packet of components with different 
asymptotic momenta.
Hence, the total wave function is also a superposition
of energy eigenstates. However, this
 wave function does not describe
everything and in particular it has not been taken into
account the wave function of the position measuring apparatus
(that one can always introduce).
It is the apparatus which is sensitive to interferences
of the different components of the wave packet, thereby giving
physical substance to the expansion coefficients, i.e. 
bringing about a localized description of the experiment,
see \cite{future2} for more details.
Clearly in cosmology such a procedure is inoperative 
since it cannot be carried out. 

\vskip .2 truecm
We wish to conclude this essay by suggesting a way to look
at  Mach's 
principle from a quantum point of view.
Mach's  principle is often stated 
``the distant stars determine our local inertial frame.'' We should
like to enlarge upon this thought so as to give it more 
dynamical content. 
Provided that first, a mean description 
in the sense of minisuperspace is meaningful and second,
the universe is macroscopic, we
have shown that one can describe 
cosmological events in terms of inertial concepts
(i.e. the cosmological time $\bar t(a)$ is used in 
the same way as inertial time).
This remarkable result, emerging from the Wheeler-DeWitt 
constraint, obtains because the phase differences of the 
WKB gravitational wave functions which occur in the description
of transitions are small,
i.e. energy differences induced by transitions are 
infinitesimal in the almost infinite sea of matter that fills
the universe.
It is, as it were, that gravity (described by WKB waves)
 possesses an enormous ``inertiality'' due to the 
enormity of the mass in the universe.
Thus we are led to enlarge upon 
Mach's principle in that use of the conventional 
notion of inertial time to describe cosmological events
requires a macroscopic universe.
\vskip .6 truecm
{\bf Acknowledgements}

\noindent
R.P. wishes to thank Serge Massar for many fruitful
discussions on these matters as well as the people of the
LPTENS of Paris wherein this work was initiated.
We both thank the theory groups of the University of Bologna
and the University of Marseille for their kind hospitality.
Our visits there permitted us to air out some of the ideas
which are expressed above.
This work received financial support from the European contract CHRXCT 920035.

\end{document}